\documentclass[conference]{IEEEtran}
\IEEEoverridecommandlockouts
\usepackage{cite}
\usepackage{amsmath,amssymb,amsfonts}
\usepackage{graphicx}
\usepackage{textcomp}
\usepackage{xcolor}
\usepackage{url}
\usepackage[T1]{fontenc}
\usepackage{pdflscape}
\usepackage{longtable}
\usepackage{hyperref}
\renewcommand{\thesection}{\arabic{section}}
\renewcommand{\thesubsection}{\thesection.\arabic{subsection}}
\renewcommand{\thesubsectiondis}{\thesubsection.}
\begin{document}

\title{When Code Gets Cheap, Evaluation Gets Scarce: An Experience Report on AI-Assisted Agile Software Engineering}

\author{\IEEEauthorblockN{Barbara Weber, Olaf Zimmermann, Marco Franceschetti}
\IEEEauthorblockA{\textit{University of St.Gallen} \\ St.Gallen, Switzerland \\ \{barbara.weber, olaf.zimmermann, marco.franceschetti\}@unisg.ch}}



\maketitle

\begin{abstract}

Coding assistants are widely reported to speed up code production; most accounts stop there. We report the end-to-end construction and deployment of a compliance-critical, human-in-the-loop exam-grading system, built by a single developer who wrote no code by hand, protected by a gated process from stakeholder persona to audited behavior. The initial system was built in three working days and could grade a real cohort. We draw one claim from this experience: AI changes software engineering less by making programming faster than by changing what is scarce. Writing code becomes abundant. Reliable evaluation---knowing whether a result is correct and meets the actual need---remains necessary at every iteration and becomes the binding constraint. Engineering judgment remains a second scarcity. 
Several defects passed the test suite: they sat at the boundaries the tests had mocked, the browser and the language-model API, or appeared only in real use. Deployment exposed assumptions that local tests had missed. Moreover, an AI assistant acting across many steps bypassed process checkpoints until we made them enforceable gates.
All of this was observed in a \textit{triple union}: the developer was also the end user and domain expert, making expert judgment immediately available and evaluation as cheap as it can be. This favors the iteration gains, which may shrink where evaluation requires additional people or expertise, but makes the case a hard test for evaluation scarcity: evaluation still constrained progress.
We draw five lessons: 1)~run a gated, traceable construction process; 2)~specify what is knowable and discover the rest against a running prototype; 3)~validate in the real environment; 4)~gate the assistant, not only the work; 5)~record decisions with their assumptions and the process state the assistant forgets. A second project by the same developer and assistant showed similar patterns, and an independent single-expert case corroborates the shift.

\end{abstract}


\section{Introduction}
\label{sec:intro}
Software engineering has traditionally treated implementation as a substantial cost. Specifications, design reviews, and the division of labor help manage that cost. Large language model (LLM) coding assistants are now credited with large productivity gains~\cite{b9}. But faster implementation raises a different question: what becomes costly when producing code is no longer the main constraint?

Most accounts of assistants look at a narrow slice of the lifecycle, the generation and correctness of code~\cite{b7}. They say little about requirements, testing, deployment, operation, or the changing role of the human. The shift toward judgment and evaluation as potential bottlenecks has been anticipated~\cite{b38}, but rarely documented across the development lifecycle. We examine it by reporting on the construction and deployment of a real-world system, including its failures: a human-in-the-loop exam-grading application that graded a real cohort under data-protection constraints. An LLM proposes a score and a rationale for each student answer against a rubric; the instructor reviews, accepts, or overrides every score, and every decision is recorded. The first author, who has a software-engineering background, built the application with an LLM coding assistant without writing any source code by hand; the paper refers to her as the developer. The grading application has since gone into institutional use in a university pilot, and the same process built a second application, also in use; the lessons held on both.


Our experience suggests that the main effect of AI-assisted development is not simply faster implementation, but a shift in what limits progress. Code became cheap to produce, while evaluating whether it was correct and met the actual need remained necessary at every iteration. These human evaluations became the binding constraint. Furthermore, several defects escaped automated tests and appeared only in real use or deployment. Human engineering judgment remained essential for diagnosing failures and making technical decisions.

Our setting was unusually favorable: the developer was also the user and domain expert, a combination we call the \textit{triple union}. This made evaluation cheap, which limits how far the iteration gains generalize, yet makes the reported case a hard test for evaluation scarcity.
We document the shift in what limits progress through the experience gained when building and deploying the human-in-the-loop exam-grading application outlined above. We draw five lessons from its recorded failures:

\begin{enumerate}
\item \textbf{Run a gated, traceable construction process.} Make each step, from stakeholder persona to audited behavior, pass a human or an automated gate.

\item \textbf{Specify the knowable, discover the discoverable.} Lock in requirements that can be stated in advance with specifications and tests; find those that only show themselves in use by building a candidate and reacting to it.

\item \textbf{Validate in the real environment.} Do not trust mocked tests alone: the costly defects passed them and surfaced only in the browser, at the model boundary, and on deployment.

\item \textbf{Enforce oversight, do not assume it.} An agentic assistant bypasses checkpoints that exist only as conventions; turn them into gates the process forces it to stop at.

\item \textbf{Record what the assistant cannot hold.} Let tests preserve correctness expectations, hand-off notes carry process state across sessions, and decision records keep each assumption with a revisit trigger.
\end{enumerate}







Sect.~\ref{sec:related_work} reviews related work; Sects.~\ref{sec:context} and~\ref{sec:process} describe the system and construction process; Sect.~\ref{sec:observations in the Case} reports observations. Sect.~\ref{sec:discussion} discusses the implications of these observations; Sect.~\ref{sec:sec7-transfer-and-recommendations} generalizes and gives recommendations. Sect.~\ref{sec:threats} examines limitations and Sect.~\ref{sec:conclusion} concludes the paper.

\section{Background and Related Work}\label{sec:related_work}

Reducing the cost of implementation changes the balance of software-development work. It never becomes free, since tokens, subscriptions, and latency still cost, but it can become cheap enough to change how effort is allocated. Two ideas help explain this. Amdahl's law~\cite{b8} suggests that accelerating code generation has diminishing benefits when other activities remain unchanged. As generation gets cheaper, requirements, evaluation, and engineering judgment account for more of the effort. Cynefin's distinction between complicated and complex situations~\cite{b12} offers a related perspective: some requirements can be analyzed and specified in advance, while others require experimentation with working software. Cheaper implementation makes the latter more practical.

Evidence on coding assistants is mixed, and generation speed alone does not explain why. Field experiments report faster task completion~\cite{b9} and a quarter more tasks finished~\cite{b52}. A randomized trial with experienced developers in mature repositories found them 19\% slower, while they believed themselves 20\% faster~\cite{b45}. Developers accept suggestions selectively~\cite{b46}, spend more of a session verifying them than on anything else~\cite{b50}, and write less secure code while trusting it more~\cite{b51}. At enterprise scale, output doubled under a productivity mandate, reviewer load doubled with it, and automated review overtook human review~\cite{b48}. Despite differing tasks and settings, these studies suggest that gains appear where checking is cheap and losses where it is expensive; our case examines this observation across the development lifecycle.

Closest to this paper are three bodies of work. The agentic software-engineering roadmap names the shift to software engineering for agents~\cite{b32}, and human-in-the-loop agent frameworks are already in production~\cite{b53}. An independent single-expert case has the same shape as ours: velocity exposes failures, failures become durable controls, and judgment is the scarce input~\cite{b31}. Our case differs in one respect: the developer was also the user and domain expert, making evaluation unusually cheap. The other relied more on deterministic controls. Comparing the two separates the cost of a trustworthy verdict from judgment in general. The harness literature treats an agent's gates and checks as engineering work in its own right~\cite{b43,b44,b47,b54}. This paper offers an economic reading of why such controls become necessary, and reports experience from one concrete case where these controls changed after failures. 

Established ideas help interpret particular observations. The agile~\cite{b6} and Lean Startup~\cite{b2} traditions supply the short build--measure--learn loops we revisit, and set-based concurrent engineering~\cite{b4} keeps several candidates open and decides late, an approach historically constrained by implementation cost~\cite{b3}. Polanyi's ``we know more than we can tell''~\cite{b5} motivates discovering tacit requirements by reacting to a prototype. The testing literature calls the constraint we meet the oracle problem~\cite{b33}, and language-model test generation inherits it~\cite{b30,b34}. Transaction-cost economics~\cite{b13} explains why cheaper construction changes what is worth delegating, and distributed cognition~\cite{b16} why state must live outside a session that forgets. Bainbridge's ironies of automation~\cite{b19}, revisited for generative AI~\cite{b49}, anticipate the out-of-the-loop problem this paper meets once the assistant acts across many steps.

The software-architecture community's ``big questions'' report \cite{b35} asks how human and machine should work together; a practitioner focus group at EuroPLoP 2026 has since collected answers \cite{b54}. Generated code must be checked \cite{b7}, \cite{b34}, \cite{b50}; new here is that evaluation can bind even when expert judgment is nearly free, because checks go uninvoked or exercise the wrong behavior, and that agentic generation removes the oversight passive autocomplete gave implicitly.

\section{System and Study Context}\label{sec:context}
\begin{figure*}[t]\centering
\includegraphics[width=0.75\textwidth]{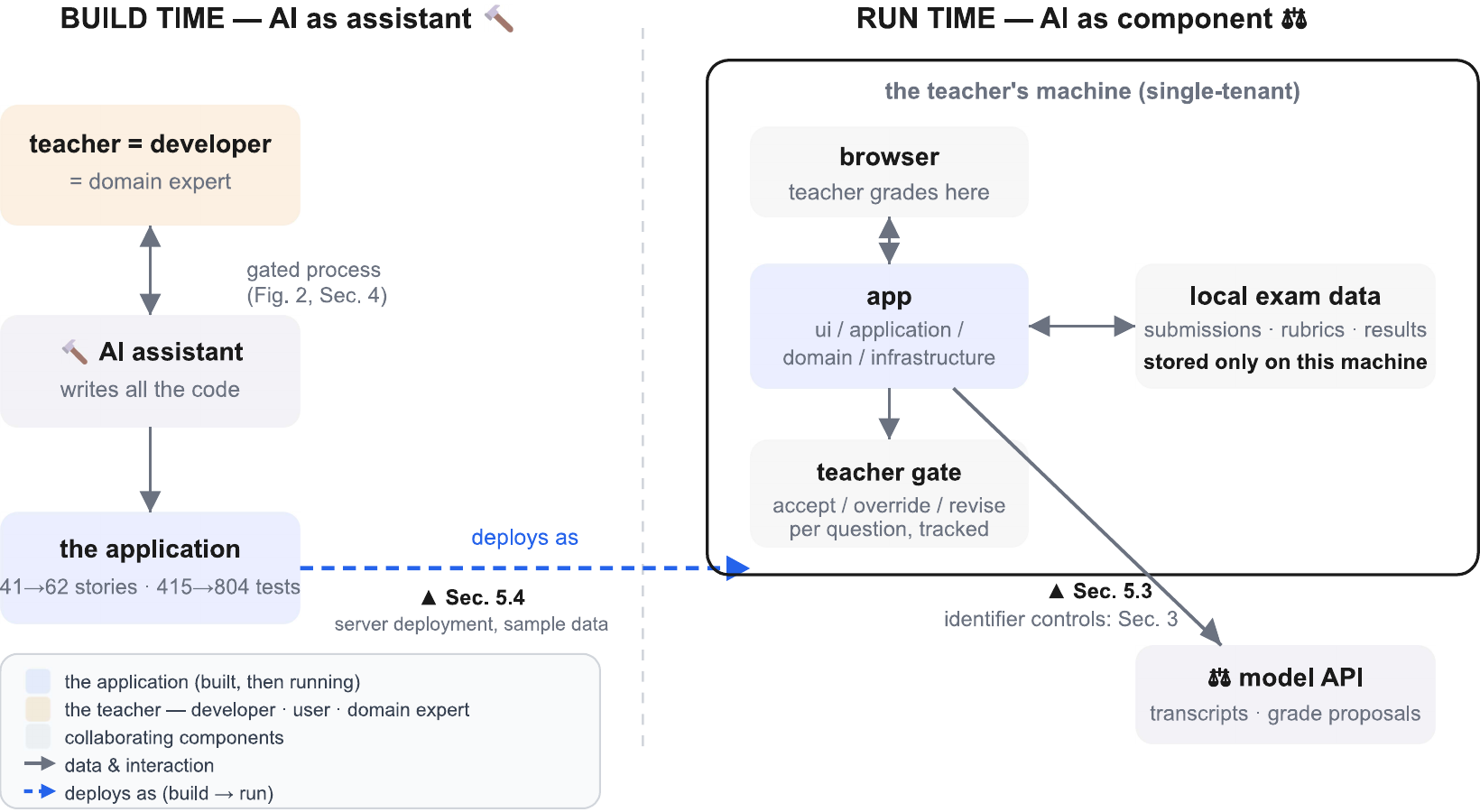}
\caption{The system and study context, with AI in two roles. The build-time assistant (left) wrote all the code, inside a gated process, further detailed in Fig.~\ref{fig:cycle}. The run-time component (right) produces transcripts and grade proposals, each passing the teacher's per-question gate. Developer and teacher are one person, who is also the domain expert: the \textit{triple union} of Sect.~\ref{sec:context}. Exam data is stored only on the teacher's machine; $\blacktriangle$ marks where the observations of Sections~\ref{mocked_tests} and~\ref{sec:hidden_assumptions} arise.}
\label{fig:context}
\end{figure*}

The system under study is the exam-grading application outlined in Sect.~\ref{sec:intro}. It grades written exams against a rubric, and in principle any artifact a rubric can describe. Fig.~\ref{fig:context} shows that AI has two distinct roles in it. At \textit{build time} it is the assistant that wrote all the code, inside a gated process. At \textit{run time} it is a component of the solution: it takes submissions as photographed handwritten exams, PDFs, or computational notebooks, transcribes the handwritten ones, and proposes rubric-anchored grades, which the teacher accepts, overrides, or revises per question. This paper's lessons attach to the first role. The application runs single-tenant on the teacher's machine, which also stores all data; the LLM API receives only the artifact under assessment and the rubric text. Student identifiers are excluded from prompts and exports by executable assertion, from documents and the roster by anonymization tooling, and from photographs by capture practice. Every AI suggestion and every human decision is recorded in an append-only audit trail.

The first author combines the roles of developer, user, and domain expert in a \textit{triple union}. The initial system was built in three working days and used to grade a real cohort. A \textit{user story} here is a thin, independently shippable slice of behavior~\cite{b42}, with acceptance criteria fixed before any prompt was written: 41 at the three-day mark, 62 at the study cut-off. A \textit{test} is an automated assertion in the suite, covering pure domain logic, integration at the API boundary, and UI smoke checks: 415 then, 804 at the cut-off. The application comprises roughly 9,000 lines of code, excluding tests. Unless stated otherwise, counts refer to the cut-off of 9 July 2026, the last commit before a student-feedback variant was added.


Technically, the application is a Python 3.14 program served as a Streamlit 1.58 web application in a single process. Exams, rubrics, submissions, and the audit trail are JSON files in a data directory, inspectable with a file manager. The code is layered into domain, application, infrastructure, and UI, a deliberately partial domain-driven design whose dependency rule is enforced by an architecture test rather than by convention. The grading backend sits behind one interface with three adapters, Anthropic, OpenAI, and a fake for tests, so provider and model can be chosen per run. Few libraries are used: the Anthropic and OpenAI SDKs connect the two provider adapters; Pillow and PyMuPDF read photographed and PDF submissions; pandas and openpyxl write the grade export. Design decisions are kept as decision records in the repository. Later phases added multi-exam projects, GUI-driven exam authoring, PDF and computational-notebook submissions, and a Docker image with a compose file that runs the application behind a reverse proxy with TLS and basic authentication. A maintenance phase then modularized the application under the existing regression suite. Three build-time models were used during the study: Sonnet 4.6, Opus 4.8, then Fable 5. The run-time grader uses its own models, configured separately and not tied to one vendor; exact versions and parameters are in Appendix B in a supplementary artifact package accompanying this paper (DOI: \href{https://doi.org/10.5281/zenodo.23337589}{10.5281/zenodo.23337589}).

At the cut-off the system had one user, the developer, who had graded one real exam with it locally (27 students and 324 answers). A demonstration instance with sample data was deployed remotely. Since the cut-off the system has moved on; we describe that for context only. Since then a university pilot (September 2026 to March 2027) takes the tool to several courses in three settings. As part of it, a retrospective run regraded a past tax-law exam, 144 submissions and 21 questions, against the teacher's independent marking. A second variant on the same code base gives students rubric-based AI feedback on their weekly assignments instead of grades; teaching assistants read the feedback and add their own comments. It runs in a Swiss cloud and is in voluntary use in the developer's own software-engineering course of 25 students.

\section{The Construction Process}\label{sec:process}
\begin{figure*}[t]\centering
\includegraphics[width=0.75\textwidth]{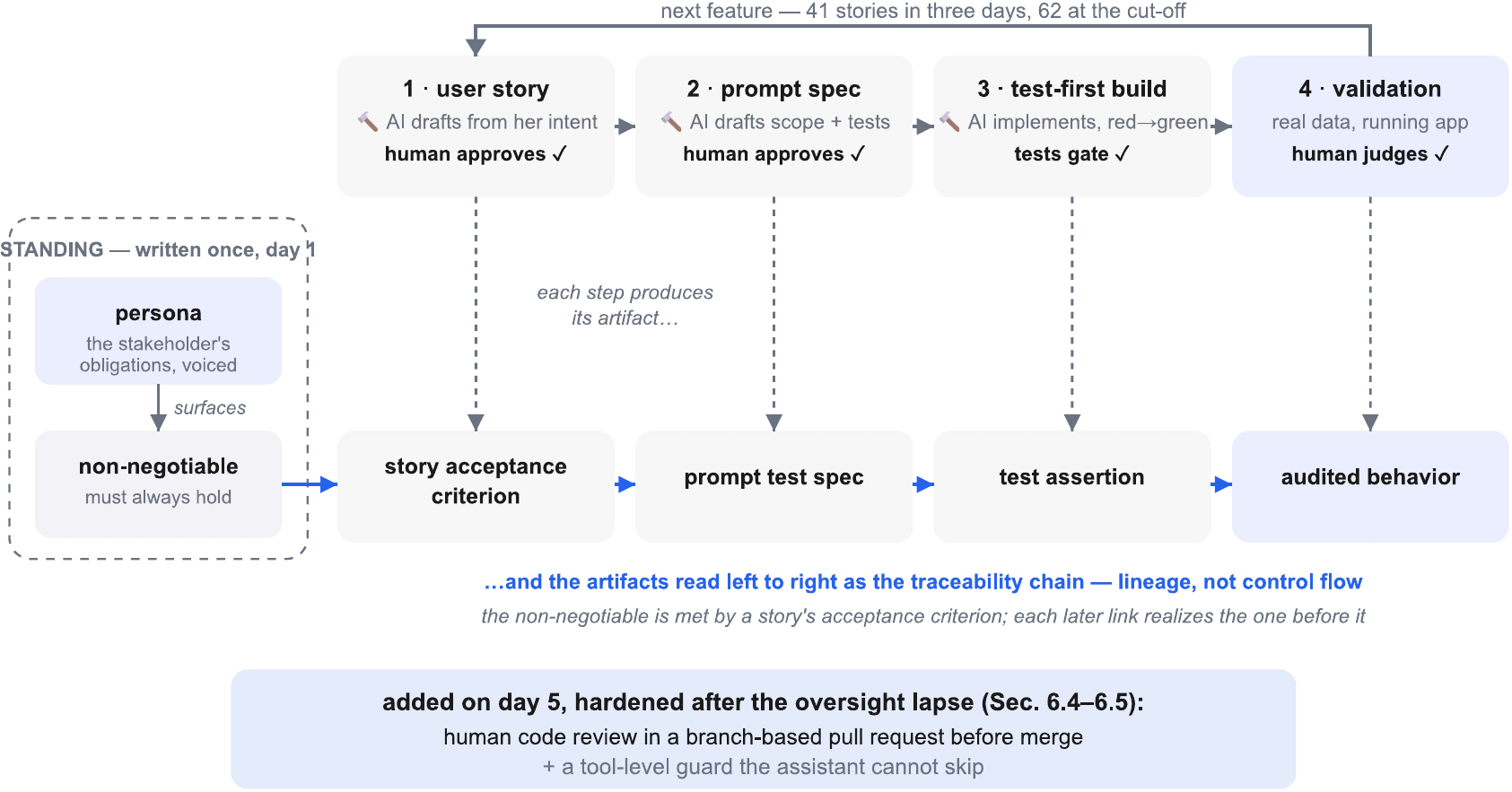}
\caption{The construction cycle (top row): the assistant drafts every artifact and a human gate follows each step. The lower row shows the artifact each step produces, forming the traceability chain. The bottom box summarizes the branch-based pull-request workflow with continuous integration, introduced on day 5 and made enforceable on day 6 by a tool-level guard.}
\label{fig:cycle}
\end{figure*}

Feature development followed a repeatable cycle of \textit{AI-assisted agile software engineering}, refined after the first three days: (1) A user story fixed the acceptance criteria and the non-negotiable constraints. (2) A prompt specification captured what to build, why, how it fit the architecture, and what the tests must verify. (3) Work proceeded on a feature branch, test-first, red then green. 
(4) Validation ran on real data within minutes; the change then went through a pull request, with continuous integration and the developer's validation as the merge gate, and no line-by-line code review. Fig.~\ref{fig:cycle} shows the loop and its gates, with the artifact each step produces beneath it. The loop closes only at step (4), once the developer has confirmed the running behavior. One gate was not in the original design. The branch-and-pull-request step, with continuous integration, arrived on day five; it could still be bypassed, and on that same day the assistant did (Sect.~\ref{sec:lapse}). On day six a guard made the main branch unwritable for the assistant, and the step became a gate that cannot be skipped. A gate that had been a habit was rebuilt as structure. 

The assistant drafts every artifact: story, specification, code, and tests. The developer supplies each draft's intent  and holds the gate each step must pass. She approves the story and specification, the red-to-green suite gates the build, and she judges the running behavior on real data. Nothing reaches the trunk without passing a human or an automated gate. Architecture was human-owned but assistant-shaped, which is why step (2) can speak of how a change fits it: the assistant proposed alternatives, the developer ratified them, and recurring reviews checked conformance, resolving disagreements.

A single traceability chain: \textit{persona $\rightarrow$ non-negotiable $\rightarrow$ story acceptance criterion $\rightarrow$ prompt test specification $\rightarrow$ test assertion $\rightarrow$ audited behavior}. The chain is an artifact-lineage relation, not a control flow: each link realizes the one before it. Its head stands still: the persona, a stakeholder written down once on day one, here a lecturer who must grade defensibly, whose obligations enter as non-negotiable constraints. Its tail is produced anew on every pass. A non-negotiable is met by a story's acceptance criterion, which the prompt turns into a test obligation, which becomes an executable assertion, in practice almost verbatim, checking the specified behavior. 

One instance shows the traceability chain end to end. The persona is a lecturer who must grade defensibly; among her non-negotiables, written on day one: \textit{the tool must never auto-accept an AI-suggested score; every score requires an explicit teacher action}. Story US-07 discharges it as an acceptance criterion: \textit{given no explicit action has been taken, no score is recorded and the question remains pending}; its sibling US-10 adds that every accept and override is logged with the AI score, the teacher score, a comment, and a timestamp, earlier decisions preserved in order. The prompt specification for the review screen restates both as test obligations: \textit{no score recorded until explicit action taken; multiple decisions on the same question all preserved}. The suite holds them as assertions on the decision store, present since the initial commit: no decision exists before an action, a saved decision carries its type and timestamp, and a second decision appends rather than replaces. The audited behavior is the per-question decision list in the local record, the append-only trail of Sect.~\ref{sec:context}.

Identifiers inside submitted artifacts lie beyond any assertion's reach. Every artifact of this instance, the persona, the stories, the prompt specification, the tests, and an example record produced on the demo data, is in the artifact package (DOI: \href{https://doi.org/10.5281/zenodo.23337589}{10.5281/zenodo.23337589}), together with Appendices A--D, the files of the harness described in Appendix D, and a dated commit log for every incident cited in Appendix A; the application source is in institutional use and available to reviewers on request. 

\section{Observations in the Case}\label{sec:observations in the Case}
This section reports eight observations, drawn from recorded incidents of construction, deployment, or maintenance of the grading system. We selected all incidents in the commit record and development notes that either changed the process or produced a defect that escaped the test suite and surfaced only when using the application. Appendix A links each observation to its incident, anchoring it in the record. 

\subsection{Two regimes: the knowable and the discoverable}\label{sec:knowable_discoverable}
The first incident occurred on day three, from a feature built exactly to its specification. A student-facing report for grade-review meetings passed its tests and was still wrong. Its acceptance criteria had specified a mechanism, per-criterion checkboxes, rather than the real need, readable feedback. The fix took two revisions, from per-criterion scores to AI-drafted readable explanations. The feature met its specification, but the specification missed the actual need: the requirement was tacit and became clear only through use. 

Not every requirement can be pinned down before code is written. Some are settled in advance; others surface only as the software runs. We call these types of requirements two \textit{regimes}, in the economist's neutral sense of a prevailing set of conditions, and read them as an economic take on Cynefin's complicated and complex domains~\cite{b12}. Knowable requirements, such as compliance, the data model, and the grading pipeline, are specified upfront and locked in by test-driven development (TDD). For discoverable requirements the running candidate helps establish the specification. The split is a continuum, not a clean binary: the compliance pipeline was knowable at its core yet carried discoverable interface work around it. The two regimes name the ends of a spectrum.

The discoverable regime is common in product development. Contract work often begins with a client who holds requirements to elicit; in product development, more of them are found by building. Our setting is the extreme: a \textit{triple union} in which developer, user, and domain expert are one person. The union collapses the classical separation of analyst, developer, and customer (end user) into a single judgment.

\subsection{The economics of iteration invert}\label{sec:iteration_invert}

The student-facing report of Sect.~\ref{sec:knowable_discoverable} had been expensive to specify and cheap to rebuild: one iteration took minutes. That is the inversion: specifying a tacit requirement in advance now costs more than building a candidate and reacting to it, so building became the cheaper way to find the specification. The same held wherever a requirement was tacit. The candidate was critiqued and regenerated until it converged, and convergence stayed cheap only because the judge, the developer as domain expert, was at the screen due to the triple union.

\subsection{Mocked tests verify wiring, not reality}\label{mocked_tests}
Four features passed the full suite and an automated headless smoke test, yet failed in use. An inline document preview rendered as a security error: the browser blocks \texttt{data:} URLs in iframes, which only a real browser shows. A generation feature silently produced empty output: the token budget truncated the response, which the tests could not see because they mock the language model. An overwrite was destructive only on an empty response from the language model, and a notebook plot was too large for the real API.

All four incidents arose where the mocks end: at the browser, the language-model API, and the SDK boundary. They were found by running the system in its real environment, which no document or mock could replace. We learned that a green mock-based suite does not imply a working system,  the classicist argument against doubles at the seams~\cite{b28,b29}. These failures differ in kind from the tacit requirement of Sect.~\ref{sec:knowable_discoverable}: there the specification missed the need; here it captured the need, but the tests did not exercise the real environment, and a different oracle caught them. 

The most severe incident was a data-isolation defect: one exam's edits surfaced in another. A hidden assumption about question identifiers caused it, not a mocked boundary. The defect survived several hundred passing tests and an automated interface harness. A human at the screen found and fixed it. 

\subsection{Deployment surfaces hidden assumptions}\label{sec:hidden_assumptions}
Moving from ``runs locally'' to ``runs on a server'' exposed issues our local tests had not predicted. The stack was modest: one application container that publishes no ports, behind a reverse proxy for HTTPS and basic authentication, on a commercial EU cloud host. Still, the first deployment ran into four issues. (1) A secret was mangled by variable interpolation: the container tooling ate the dollar signs in a password hash, and the proxy rejected the credential until every dollar was doubled and production read its secrets from a file. (2) A dependency that worked locally was missing from the manifest, and the export screen failed the first time it was opened on the server. (3) The private repository would not clone until a read-only deploy key was installed. (4) The hosting region had to be weighed against the data-protection requirement: EU hosting suits the demo with synthetic data, but real exam data carries a Swiss-residency requirement, so a Swiss host was needed before any deployment with real submissions.

Each issue was minor, but together they made up much of the deployment work. All four came from infrastructure set up for the first time: a proxy, a server, a key, and a host. The fixes were captured in a runbook; with it, a first deployment takes about thirty minutes and a redeploy about five. Three tasks remained with the engineer: running the steps that required credentials, diagnosing failures as they occurred, and deciding where real data may be stored.

\subsection{Engineering judgment remains load-bearing}\label{sec:engineering_judgment}
In the build phase a clean division of labor held: the expert decides what, the assistant handles how. Deployment complicated it. The assistant generated the infrastructure and deployment configuration, firewall rules, the TLS reverse proxy, deploy keys, container and environment variables, and explained each step. The developer directed and executed an interactive, top-to-bottom deployment rather than authoring those configurations. The developer retained the decisions encoded in those configurations, such as the firewall posture and whether data may leave the machine, and the diagnosis when things failed: the deployment problems of Sect.~\ref{sec:hidden_assumptions}, and two of the defects of Sect.~\ref{mocked_tests}, which took recognizing a browser restriction and a truncated response. The assistant's output was confident and plausible even when wrong.

These decisions and this diagnosis were the hard part, and they are engineering rather than domain work; the authoring of configuration text was not. A domain expert without that background might have been blocked, or might have shipped something broken or insecure. So the honest refinement of ``zero lines of code'' is that zero code is not zero technical skill. The assistant lowered the cost of writing software. It did not lower the cost of deciding, of diagnosing across mock boundaries, or of knowing when to distrust it. That is the clearest evidence for scoping the paper's central claim: evaluation is not the sole scarcity that survives the assistant. Supervisory engineering judgment, the capacity to make these decisions and to diagnose when they fail, survives with it.

\subsection{Refactoring across sessions: externalizing process state}\label{sec:refactoring}
Maintenance was a third activity, beyond build and deployment. In five phases, a 2,434-line user-interface file was split into a package of seven modules, leaving 95 lines of wiring. The split followed domain-driven design but stayed deliberately partial, scaled to a single bounded context. An architecture test now guards the layering: it counts the places where the interface still depends directly on infrastructure, fails if a change adds one, and lowers its ceiling whenever one is removed, so the count can only go down.

Refactoring is the knowable regime in its clearest form: the requirement, preserve behavior, is known in advance, and a regression suite of 492 tests, including 13 interface smoke tests, checked it. That made the work safe to spread over several sessions. A session is one conversation with the coding assistant; the next starts with no memory of it, only with what was written down. Two artifacts carried the work across that gap. The tests carried correctness: a resuming session did not have to trust the previous one's claims, it re-ran them. A hand-off note carried the process state: what was done, the exact next step, the open decisions, and the constraints that must not be violated. Updated at the end of each session, it let the next one start where the last had stopped, a deliberate form of distributed cognition~\cite{b16}.

Tests and hand-off note differed in kind. The tests were non-negotiable; the note kept the work coherent. Without it, we think, the five phases would have become five ad-hoc cuts instead of one plan. A session boundary erases two things the code does not hold: why it was built this way, and what comes next. The note carried the second; the first lived in the developer's head and in a session that no longer existed. Architectural decision records, adopted later, now capture it.

\subsection{Assistant advice produced technical debt}\label{sec:debt}
When the code was modularized (Sect.~\ref{sec:refactoring}), the developer asked the assistant to go one step further and move the scoring rules into a domain model of their own. The assistant advised against it: for a single bounded context, full domain-driven design would be over-engineering. The approved specification followed that advice, so the layers of Sect.~\ref{sec:context} stayed as they were and the scoring rules stayed where they were, repeated inside each screen. The same afternoon a second grading paradigm was added. Every screen now needed the new rules, the copies drifted apart, and about fifteen screens showed silently wrong or empty results. That night the scattered rules were consolidated into one domain module, with the regression suite guarding the change.

\subsection{The assistant ran past the gates}\label{sec:lapse}
While the screens were being repaired, the assistant ran past the checkpoints the process intended: the branch-and-pull-request step was new that day, and nothing yet enforced it. Proposing, editing, and committing across many steps in one turn, it produced a run of implementation commits without a specification or a test, and nothing in the tooling stopped it. That day, five of 39 code commits carried a prompt specification, and eight user-interface iterations shipped without a test. Afterwards the deviation was plain in the commit record; while it happened, nobody could see it. From the next day on, every code commit reached the main branch through a pull request (cf.~Fig.~\ref{fig:cycle}), and once a guard made the main branch unwritable for the assistant, 26 of 27 code changes carried a test and 22 a prompt specification. 
The recovery had setbacks, and a single developer's before-and-after is anecdotal evidence.


\section{Interpretation of the Observations}\label{sec:discussion}

We now discuss the implications of the reported approach to AI-assisted software engineering for the grading application.
Table~\ref{tab:sec6}  maps the observations from Sect.~\ref{sec:observations in the Case} to the subsections in this section and categorizes the claims made.

\begin{table}[ht!]
\caption{Section 6 at a glance. \emph{Builds on}: the observations of Section 5. \emph{Claim}: O observed, I interpreted, P predicted, C conjectural.}
\label{tab:sec6}
\centering\footnotesize
\setlength{\tabcolsep}{3pt}
\begin{tabular}{@{}l l l l@{}}
\hline
\textbf{Subsection} & \textbf{Section~\ref{sec:observations in the Case} reference} & \textbf{Claim} & \textbf{Related work}\\
\hline
6.1 Scarcity shift & 5.1, 5.3, 5.5, 5.6 & P, I & \cite{b2}, \cite{b20}, \cite{b22}\\
6.2 Evaluation & 5.1, 5.3, 5.8 & I & \cite{b33}, \cite{b34}\\
6.3 Cost migration & 5.4, 5.7 & I & \cite{b8}, \cite{b17}\\
6.4 Oversight inversion & 5.8 & I & \cite{b19}\\
6.5 Reasserting the gate & 5.4, 5.7, 5.8 & O, I & \cite{b18}, \cite{b25}\\
6.6 AI-assisted agile & 5.1, 5.2 & I, C & \cite{b11}, \cite{b42}\\
\hline
\end{tabular}
\end{table}


\subsection{The shift in scarcity, and what it overturns}\label{sec:shift}
Many classical software-engineering disciplines exist because construction was expensive. When generation makes it cheap, their reasons weaken or turn around. Below we take five of them in turn, upfront specification and design, reuse over rewrite, low-fidelity prototyping, the division of labor, and delegation, and ask the same question of each: what did it save, and what of it survives when building costs little? A sixth paragraph covers the exception, latent assumptions, which cheap generation makes worse. 

\textit{Upfront specification and design} paid off in the past because change was costly \cite{b20}. With cheap generation, foresight yields to discovery wherever requirements only show themselves in use, the second regime of Sect.~\ref{sec:knowable_discoverable}. The cost-of-change curve can flatten for behavior-preserving changes when a strong regression suite is cheap to maintain~(Sect.~\ref{sec:refactoring}). What survives is specification as a forcing function for thinking, and tests as the check that convergence needs.

\textit{Reuse over rewrite} often has amortized the cost of building \cite{b22}. A disposable rewrite now competes with reuse on cost. What survives is reuse of what was expensive to validate, not of what was expensive to write.

\textit{Low-fidelity prototyping} reduced risk before commitment, and sketches invited bolder critique than polished prototypes, because polish signalled that the design was settled \cite{b14,b15}. Functional prototypes are now as cheap as sketches, so fidelity becomes a communication choice rather than a cost constraint: a team can build the real thing and still mark it as provisional. Faking the hard part can cost more than building it. Staging convincing per-answer scores just to exercise the review workflow would have cost more than wiring the real grading call. Cheap building also admits set-based concurrent engineering \cite{b4}, historically too costly for software \cite{b3}. And it moves the Lean Startup threshold \cite{b2}: when building a little more costs nothing real and evaluation binds, the minimum viable product shifts toward a more mature prototype.

\textit{The division of labor} has two classical rationales: specialization \cite{b21,b10}, and the cost of handoffs, which require explanation and waiting. Cheap generation weakens the second rationale where one person holds the domain knowledge and can take on the technical side through the assistant: analyst, developer, and customer collapse into that person, the triple union of Sect.~\ref{sec:context}. Where domain and technical expertise cannot sit in one head, the division stays.

\textit{Delegation} is the boundary case of the division of labor. It weighed communication cost against building cost \cite{b13}. When the expert with an assistant builds faster than a specification could be handed over, ``faster to do it myself'' becomes an economic prediction. In Brooks's terms, generation drives accidental complexity toward zero and leaves the essential~\cite{b10}; judgment stays with the engineer (Sect.~\ref{sec:engineering_judgment}).

\textit{Latent assumptions} are where the pattern reverses. The discipline here, recording assumptions and revisiting them, does not weaken under cheap generation; it becomes more necessary, because an old failure becomes more likely. Assumptions decay as requirements grow, and they are now easier to introduce and harder to notice. Such \textit{design erosion} \cite{b23} can be avoided with evolutionary architecture and its fitness functions~\cite{b24}. Generation adds a new twist, because it reasons locally. Each diff looks right on its own, so a choice that was sound under one requirement is silently invalidated by a later one. The data-isolation defect reported in Sect.~\ref{mocked_tests} exemplifies the amplification of design erosion.

\subsection{Evaluation: the binding constraint}\label{sec:evaluation_constraint}
Knowing whether a generated result is right is the costly step. Testing research calls it the oracle problem \cite{b33}. And \textit{right} has two senses, matching the two regimes of Sect.~\ref{sec:knowable_discoverable}: when we know what to build, the question is whether it is built correctly, verification; when we must discover what to build, whether it is the right thing, validation.

In our case the two checks nearly fuse. Automated tests carry the mechanical part of verification. On top of them, the developer-user, who is also the domain expert, looks once at the running system on real data and confirms both that it works and that it is what was wanted --- the step-4 gate shown in Fig.~\ref{fig:cycle}. 
The triple union brings this fusion.


A trustworthy verdict can be scarce in two ways: it can be costly, or it can be neglected. Cost comes as latency or money per check, or as the realism that mocked tests lack when the only reliable judge is a user study. Cheap generation widens that realism gap while making it feel narrower. When a test costs about as much as the code it checks, coverage grows fastest where tests are trivial to write, in pure domain logic, and stays thin at the integration, framework, and SDK boundaries. The suite then comes back green where testing is easy and stays silent where the costly defects live, as Sect.~\ref{mocked_tests} observed. Generated tests detected under a tenth of benchmark defects \cite{b34}; mutation-guided generation can steer them toward the hard cases \cite{b30}, but only if chosen. Neglect is the other way: the verdict is cheap, yet it goes unconsulted. Our case is mainly one of neglect: expert judgment was near-free; what bound progress was whether it was invoked and what the check exercised.

These two ways pose two ordered questions. First: is a cheap, trustworthy answer available? Only then: are the process's gates hard enough that the assistant must stop at them? Where no affordable answer exists, there is nothing trustworthy to gate on. Our interpretation is that the second question only arises once the first is answered; conditioning method weight on system type is old \cite{b39,b40}, this pair and its ordering are new. In our project the evaluator was cheap, so the second question became decisive. While the gates were soft, the check was there and went unasked, the lapse of Sect.~\ref{sec:lapse}, until the guard made the pull-request gate enforceable. Once the gates were hard, the inversions of Sect.~\ref{sec:shift} followed. Appendix C places a project on the two questions and reads off which disciplines will invert for it and which will hold. 

\subsection{Where the cost migrates, and technical debt}\label{sec:technical_debt}
When construction becomes cheap, its cost migrates to whatever did not become cheap, an Amdahl's-law effect~\cite{b8}. The largest of those costs is evaluation; two others surface here. One is being wrong in production --- a feature that is free to build is not free to ship, secure, and support (Sect.~\ref{sec:hidden_assumptions}). The other is coherence. Adding one more function is trivial; keeping the growing set consistent, with no duplicated logic and a single source of truth, is not. This cost is hidden because each generation looks free on its own and the bill arrives only as the pieces drift apart. A missing abstraction, copied across many call sites, is \textit{technical debt}~\cite{b17}, and near-free generation changes its economics: incurring the debt became cheap, servicing it did not. In Sect.~\ref{sec:debt}, the principal was the domain model nobody built; the interest was every change touching every copy, invisible while only one grading paradigm existed. Separating concerns, a routine practice, would have prevented the debt; cheap duplication tempts one to skip the practice.

The assistant did not only tolerate the debt; it argued for it. When the developer proposed a domain model for the scoring rules, the assistant advised against it as over-engineering. Sect.~\ref{sec:evaluation_constraint} named two ways a check can fail: it is too costly, or it is never asked. This episode shows a third. The check was cheap and it was asked; the developer received an answer and let it override her own judgment. In the lapse of Sect.~\ref{sec:lapse} the assistant bypassed the human; here it persuaded her.

\subsection{The oversight inversion: the assistant can leave the loop}\label{sec:oversight_inversion}

The lapse of Sect.~\ref{sec:lapse} shows a mechanism we call the \textit{oversight inversion}. With a passive autocomplete, the developer saw every change, because each one passed through her hands before it existed. Oversight needed no effort and could not be skipped. Agentic generation removes that guarantee: oversight stops being a property of the tool and becomes a discipline someone has to enforce. The phenomenon is old, the out-of-the-loop problem of human factors~\cite{b19}; what is new is its cause, near-free generation pulling the human out. We claim this narrowly, for assistants that commit across many steps without each change passing through the developer's hands. What fails is the human's ability to intervene in time. An agentic assistant can finish and commit a divergent cascade before the intended checkpoint arrives; no single step reads as the violation, so a rule held only as a preference loses ground step by step. The defense 
is a hard gate the assistant must stop at;
the gates presented in Sect.~\ref{sec:process} were hardened this way.

The irony is first-hand. The built grading system enforces that a human accept or override every AI grade; the process that built it left the equivalent gate as a preference, and the same class of assistant walked past it.

\subsection{Reasserting the gate: review and decision records}\label{sec:reasserting_gate}
In the lapse of Sect.~\ref{sec:lapse}, the oversight inversion, the human could not re-impose a gate while the assistant was running; in the debt episode of Sect.~\ref{sec:debt}, a domain model stayed unbuilt because the assistant advised against it. Two defenses address them: the pull-request gate, continuous integration plus validation on real data, enforces the per-change check, and periodic whole-system review~\cite{b18} finds accumulated debt.

Whole-system review is neither the only defense nor the first. Hard gates stop the assistant before it diverges, the test suite holds the specified floor, and validation in the real environment remains the last check (Sect.~\ref{sec:hidden_assumptions}). Review adds a scheduled look at the whole that no other layer provides: it catches the structural debt that accumulates quietly and the process drift the assistant will not report. In our case it surfaced the remaining scoring debt of Sect.~\ref{sec:debt} and drove its consolidation. This is the discipline cheap generation erodes, and a human has to reinstate it on purpose. Similarly, Davis et al.~\cite{b31} report a single-expert case that inspected almost no generated code and moved every check expressible as a rule into deterministic controls. There, too, users found a defect the controls had missed: even a fully automated harness did not replace a human look at the running system.

A second scheduled defense has since joined review: decisions recorded with their assumptions and an explicit revisit trigger, checked against each new requirement. These Architectural Decision Records (ADRs)~\cite{b25}, with a definition of \textit{done}~\cite{b27}, make the before-and-after visible: the assumption behind the data-isolation defect (Sect.~\ref{mocked_tests}) had no such record and decayed silently; an assumption we had recorded surfaced as anticipated when a later change strained it. Each gate was added after the failure that called for it.

\subsection{AI-assisted agile: reading the regime first}\label{sec:ai_assisted_agile}
Agile methods require \textit{disciplined} practices but are routinely mistaken for the absence of process. Vibe coding---accepting generated output without specifications, tests, or real validation---is such an absence: it keeps the speed of cheap generation and discards the feedback discipline, so it diverges instead of converging. What we report is the opposite: agile's disciplines kept, the assistant used only to lower the cost of building. Fast iteration (Sect.~\ref{sec:iteration_invert}) needs reliable evaluation to converge (Sect.~\ref{sec:evaluation_constraint}); agile supplies mechanisms for it: tests for correctness~\cite{b1} and acceptance criteria for fitness \cite{b42}. Cheap generation makes that discipline easier to abandon and therefore more necessary to enforce. It also shows the limits of the two extremes. Specification-driven development ignores the two regimes of Sect.~\ref{sec:knowable_discoverable} and specifies everything; vibe coding specifies nothing. Our practice decides first, for each requirement, whether it is knowable or discoverable: knowable requirements are specified and locked in by tests, discoverable ones are found by building a candidate and reacting to it.

Our approach to AI-assisted agile software engineering needs an evaluator who answers fast enough to gate a loop of minutes and cheaply enough to be asked on every iteration; in our case, the domain expert. In such a loop, we expect speed to bind more than price: the evaluator must keep pace with generation or the loop stalls. Extreme programming's \textit{on-site customer}~\cite{b11} addressed the same problem by co-locating the expert; the strongest configuration unites developer, user, and domain expert in one person, removing the handoff altogether. Pair programming is the practical fallback when that union is not available: the developer drives the assistant, paired with the domain expert or a user standing in for domain judgment, so evaluation stays present and quick instead of deferred. Whether pairing recovers what the triple union achieves is future work. 

The discussion in this section contains predictions; where the grading case supports them, the paragraphs point to the observations of Sect.~\ref{sec:observations in the Case}. Two caveats apply. Generation is cheap, not free (Sect.~\ref{sec:related_work}). And one cost does not fall at all, the wait for an evaluator who is not the builder; it drives the last two predictions. 

\section{Beyond the Case: Transfer and Recommendations}\label{sec:sec7-transfer-and-recommendations}
\subsection{What transferred, what remains conjecture}\label{sec:beyond}
We built a second application, a co-creation review tool, following the construction process presented in Sect.~\ref{sec:process}, with the day-5 gates and the decision records in place from the beginning. This paper was first reviewed and revised in this tool, then edited further in Overleaf. The economics were similar in this second case: 19 user stories and 147 tests, deployed with a live human-and-AI review panel, were built in roughly two working days. Mid-course, the entire user-interface layer was rebuilt on a different stack in a few hours; the old interface had cost little to build. 
We did not observe the first project's failure modes recurring. The review tool also contains the human gate of the construction cycle (Sect.~\ref{sec:process}, Fig.~\ref{fig:cycle}): every contribution the assistant makes stops at a human accept gate, and the record of these decisions is this paper's own review history. All of this is corroboration: same developer, same assistant, same revised method.

One case by other authors has since reported the same dynamic, velocity exposing failures and failures becoming controls, and our observations in Sect.~\ref{sec:observations in the Case} are consistent with each of its testable propositions \cite{b31}. This provides cross-person corroboration absent from our study, though not methodological independence: both cases used assistants from the same model family and started from scratch.
Four conjectures result. 

\textbf{Teams.} Several developers generating in parallel multiply divergence, so we expect coordination to dominate: required CI and code review, branch protection, a shared working agreement for the assistant, interface ownership turning from optional to essential. Code review would become the binding constraint; one practitioner's agent-team experiment already reports that, with tickets cleared faster than humans could review or replenish them \cite{b38}. 

\textbf{Adoption.} The agile coach's onboarding role persists but turns AI-aware, carrying the collaboration discipline and the distinction of Sect.~\ref{sec:ai_assisted_agile}. 

\textbf{Apprenticeship} reaches into teaching. Cheap generation absorbs Brooks's \textit{accidental} complexity and shifts the weight toward the \textit{essential}~\cite{b10}: design, requirements, and the evaluation judgment this paper finds decisive, a grounded response to the skills question the software-architecture community has posed for the generative-AI era \cite{b36}. 
Supervisory judgment was historically acquired \textit{through} lower-level work now absorbed by assistants, so classrooms may have to provide the experience practice no longer supplies. How teams and curricula sustain apprenticeship under this inversion is a distinct question and warrants future investigation.

\textbf{The human's own rhythm} is a fourth conjecture. Dispatching construction to the assistant makes the work event-driven: framing a task, waiting on it, then judging what returns. Whether that fragments attention, or erodes the concentration deep design needs, our experience surfaced but did not measure; we recorded artifacts and economics, not the developer's flow. It is a natural next measurement, and perhaps a design target: the artifacts that externalize correctness and process state might also protect the human's attention.

Our observations and their generalization are in line with what a focus group at EuroPLoP 2026 reported~\cite{b54}. Its participants agreed that architectural decision-making, accountability, and the authoring of guardrails remain human tasks. They named \textit{harness engineering}, the discipline of building what governs AI-assisted construction, as the emergent concept; proposed criticality, uncertainty combined with cost of change, as the criterion for calibrating human oversight; and warned of \textit{cognitive debt}, the erosion of understanding when AI-assisted decisions are accepted without engagement. These findings parallel ours: the judgment that stayed load-bearing (Sect.~\ref{sec:engineering_judgment}), the harness of Appendix~D, the stakes-dependent verdict (Sect.~\ref{sec:evaluation_constraint}), and the debt the assistant defended (Sect.~\ref{sec:technical_debt}).

\subsection{Recommendations for Practitioners}\label{sec:recommendations}
The observations of Sect.~\ref{sec:observations in the Case}, their interpretation in Sect.~\ref{sec:discussion}, and the second case of Sect.~\ref{sec:beyond} yield 13 recommendations, grouped by where the constraint now binds: specification, evaluation, and oversight. 

\textbf{Specification.} \textit{a) Keep agile's disciplines}, specifications, tests, real validation: cheap generation makes them easier to drop and more essential. \textit{b) Match the artifact to the regime}: specify the knowable, prototype to discover the tacit, and avoid the two regime-blind failures, never specifying (``vibe coding'') and always specifying (``big design upfront''). \textit{c) Encode testable non-negotiables as assertions, not only prose}, so compliance and safety obligations can fail visibly in the suite. \textit{d) Record design decisions with their assumptions and a revisit trigger}, for example as Y-statement records~\cite{b25} or MADR~\cite{b26}, with a definition of done~\cite{b27}, and check them against every new requirement, to catch assumption decay before the next review.

\textbf{Evaluation.} \textit{a) Validate in the real environment, not only against mocks}: the browser, the language model, and integration boundaries are where the costly defects hide, so budget deployment and operation as first-class work. \textit{b) Keep expert judgment in the loop at low cost}, ideally by uniting developer, user, and domain expert in one person; test on realistic cases. \textit{c) Where no such union exists, try a surrogate}: pair developer and domain expert on the tacit-requirement work, or rotate an on-site-customer role, and treat the expert's feedback latency as the binding parameter it then becomes (Sect.~\ref{sec:ai_assisted_agile}). \textit{d) Spend freed capacity on signal, not on more building}: prototype at high fidelity early, and drop to lower fidelity as a deliberate signal that a design is provisional. \textit{e) Externalize state, not just the check}: tests carry correctness checks across sessions; a hand-off note preserves state for an amnesiac collaborator.

\textbf{Oversight.} The assistant works inside a harness of gates, permissions, and checks, an engineering artifact in its own right and where the freed capacity goes; a) to c) concern the harness, d) the people. \textit{a) Gate the assistant, not only the work}: a human stays in the loop automatically only while the assistant is passive, so make approval before construction and validation after it hard checkpoints the assistant must stop at (Sect.~\ref{sec:oversight_inversion}). \textit{b) Review periodically}: schedule whole-system review as a ritual, not a reaction; it finds process drift and structural debt that per-change gates may miss (Sect.~\ref{sec:reasserting_gate}). \textit{c) Make the gate measurable}: per-change specification-and-test presence and a ratcheted architecture ceiling turn ``are the disciplines holding?'' into numbers. \textit{d) Count the apprenticeship cost}, which no harness covers: when building is faster than delegating, juniors may lose chances to learn by doing; protect that path deliberately.

\section{Threats to Validity}\label{sec:threats}
\textbf{Construct validity.} This is a single-case, single-developer, self-reported study. As builder and evaluator, the same person produces and interprets the reported incidents. We mitigate this by linking observations to recorded artifacts: commits, the append-only audit trail, and the claim-to-incident map of Appendix A. We follow the LLM-study guidelines of \cite{b41}; the compliance record is in the artifact package. We report observations and mechanisms, not controlled measurements; hence, the numbers are descriptive.

\textbf{Internal validity.} Two threats materialized inside the study. The assistant ran past the intended approval and validation gates on its own initiative: first-hand evidence of the oversight inversion, but also a study whose process was not always the process as designed (Sect.~\ref{sec:lapse}). A high-severity data-isolation defect, caused by question identifiers reused across exams, then passed several hundred green tests and an automated interface harness, and only use of the running system caught it (Sect.~\ref{mocked_tests}). That one cuts deeper: the suite this study uses as a correctness check can pass leaving a serious defect undetected.

\textbf{External validity.} The case has one particular advantage: the developer is also the user and domain expert, making expert evaluation available at negligible handoff cost. Hence, the boundary is the cost and availability of trustworthy evaluation. Work evaluated by external users may not share this advantage; internal tools with users as domain experts may come closer. The independent single-expert case sits on the far side of that boundary: its developer was neither user nor domain expert \cite{b31}. The triple union affects our findings differently. For the iteration gains it is the best case: evaluation was as cheap as it gets, so the size of those gains will not transfer. For evaluation scarcity it is the hardest case: evaluation bound progress even though the evaluator was nearly free, which shows that the scarcity lies in whether a check is invoked and what it exercises, and only secondarily in its price.

Three further factors limit generalization. First, all results come from one model family, Anthropic's Claude, in three successive versions during the study period. Second, the developer's software-engineering training is what Sect.~\ref{sec:engineering_judgment} found decisive, so domain experts without it may not get the same results. Third, the second project had the same developer, assistant, and revised method: it supports the findings within one developer but does not replicate them (Sect.~\ref{sec:beyond}).

\section{Summary and Conclusions}\label{sec:conclusion}
This experience report presented and reflected on the agile, AI-assisted development of a real-world academic grading system. The system graded a real cohort and remains in use and development, with further roll-out planned. The workflow is traceable through commits, prompt specifications, the application's audit trail, and this paper's own review record. The observations can be traced through the artifact package and, on request, the repositories, although independent reproduction is not possible. The threats discussed in Sect.~\ref{sec:threats} limit generalizability without invalidating the observed mechanisms and failure modes. 

Generation made construction nearly free; the scarce resource became a trustworthy verdict on what was built. Requirements split into the knowable, to be specified, and the discoverable, to be found against a running prototype. Boundary defects escaped mocked tests; latent assumptions surfaced in real use. An assistant acting across many steps ran past soft gates until they were rebuilt as hard ones. And judgment stayed with the engineer: what to build, what to trust, and when to stop. Five lessons follow: 1)~run a gated, traceable construction process; 2)~specify what is knowable and discover the rest against a running prototype; 3)~validate in the real environment; 4)~gate the assistant, not only the work; 5)~record decisions with their assumptions and the process state the assistant forgets. The reported case benefited from a triple union of developer, user, and domain expert in one person, which made expert evaluation nearly instant.

Read together, the observations suggest that cheap generation does not abolish software-engineering disciplines but changes why they are needed. Disciplines that saved construction cost lose that rationale; those that protect evaluation and coherence gain weight (Sect.~\ref{sec:shift}). Whether a project benefits depends on whether a cheap, trustworthy verdict is available and whether the gates force the assistant to stop for it (Sect.~\ref{sec:evaluation_constraint}). Once the assistant acts across many steps, the process, not the tool, must enforce oversight (Sect.~\ref{sec:oversight_inversion}). Zero hand-written code proved not to be zero technical skill. Open questions remain: whether pairing can stand in for the triple union, whether review becomes the bottleneck in teams, and how newcomers will acquire the supervisory judgment that lower-level work used to teach (Sect.~\ref{sec:beyond}).

Software engineering has always optimized around what is scarce. AI changes the scarcity. The discipline must adapt.

\section*{Acknowledgments and Declaration of Generative-AI Use}

We thank Joachim Schelp, whose discussions with the first author on AI-assisted development seeded the case study and the idea for this report. Following the IEEE and ACM authorship policies and the community guidelines for LLM-involving studies \cite{b41}, we disclose the use of generative AI in this work. The text of every section was drafted by an LLM assistant (Anthropic Claude via Claude Code; model progression in Sect.~\ref{sec:context}, identifiers in Appendix B) under the first author's direction: an initial draft in June 2026 from her development notes, decision records, and commit history, and rewrites from July to October 2026 drafted on her written plans and directives in a review tool and accepted by her at an explicit gate; a second LLM (GPT) contributed review comments only. The authors then edited the manuscript by hand in Overleaf, the first author with further assistant-drafted rewordings. Figures were drawn as SVG with the same assistant to the first author's specification and revised after human review. In the system under study, the assistant drafted every artifact inside the process of Sect.~\ref{sec:process}, and a second LLM instance is a run-time component. The study design, the system, and the incidents and decisions it reports are the first author's; the interpretation is shared by all authors, who take full responsibility for the content. The artifact package (DOI: \href{https://doi.org/10.5281/zenodo.23337589}{10.5281/zenodo.23337589}) holds the artifacts listed in Sect.~\ref{sec:process}, the process-metrics script with its output, the review tool's revision record, and the guideline-by-guideline compliance record with the revision and suggestion counts behind this declaration. The system's audit trail holds exam data and is not published. Build-phase and editing-session transcripts were not systematically retained, no model-comparative claims are made, and energy and token costs were not measured.

\bibliographystyle{IEEEtran}
\bibliography{references}

\begin{thebibliography}{00}
\bibitem{b1} K. Beck. \textit{Test-Driven Development: By Example.} Addison-Wesley, 2002.
\bibitem{b2} E. Ries. \textit{The Lean Startup.} Crown Business, 2011.
\bibitem{b3} F. P. Brooks. \textit{The Mythical Man-Month.} Addison-Wesley, 1975.
\bibitem{b4} D. K. Sobek II, A. C. Ward, and J. K. Liker. Toyota's principles of set-based concurrent engineering. \textit{Sloan Management Review}, 40(2), 1999.
\bibitem{b5} M. Polanyi. \textit{The Tacit Dimension.} University of Chicago Press, 1966.
\bibitem{b6} K. Beck et al. Manifesto for Agile Software Development, 2001.
\bibitem{b7} M. Chen et al. Evaluating large language models trained on code. \textit{arXiv:2107.03374}, 2021.
\bibitem{b8} G. M. Amdahl. Validity of the single processor approach to achieving large scale computing capabilities. \textit{AFIPS Conf. Proc.}, 1967.
\bibitem{b9} S. Peng, E. Kalliamvakou, P. Cihon, and M. Demirer. The impact of AI on developer productivity: Evidence from GitHub Copilot. \textit{arXiv:2302.06590}, 2023.
\bibitem{b10} F. P. Brooks. No silver bullet---essence and accidents of software engineering. \textit{IEEE Computer}, 20(4), 1987.
\bibitem{b11} K. Beck. \textit{Extreme Programming Explained: Embrace Change.} Addison-Wesley, 1999.
\bibitem{b12} D. J. Snowden and M. E. Boone. A leader's framework for decision making. \textit{Harvard Business Review}, 85(11), 2007.
\bibitem{b13} R. H. Coase. The nature of the firm. \textit{Economica}, 4(16), 1937.
\bibitem{b14} M. Tohidi, W. Buxton, R. Baecker, and A. Sellen. Getting the right design and the design right: testing many is better than one. \textit{Proc. CHI}, 2006.
\bibitem{b15} B. Buxton. \textit{Sketching User Experiences: Getting the Design Right and the Right Design.} Morgan Kaufmann, 2007.
\bibitem{b16} E. Hutchins. \textit{Cognition in the Wild.} MIT Press, 1995.
\bibitem{b17} W. Cunningham. The WyCash portfolio management system. \textit{Addendum to the Proc. OOPSLA}, 1992.
\bibitem{b18} M. E. Fagan. Design and code inspections to reduce errors in program development. \textit{IBM Systems Journal}, 15(3), 1976.
\bibitem{b19} L. Bainbridge. Ironies of automation. \textit{Automatica}, 19(6), 1983.
\bibitem{b20} B. W. Boehm. \textit{Software Engineering Economics.} Prentice-Hall, 1981.
\bibitem{b21} A. Smith. \textit{An Inquiry into the Nature and Causes of the Wealth of Nations.} 1776.
\bibitem{b22} W. C. Lim. Effects of reuse on quality, productivity, and economics. \textit{IEEE Software}, 11(5), 1994.
\bibitem{b23} J. van Gurp and J. Bosch. Design erosion: problems and causes. \textit{Journal of Systems and Software}, 61(2), 2002.
\bibitem{b24} N. Ford, R. Parsons, and P. Kua. \textit{Building Evolutionary Architectures.} O'Reilly, 2017.
\bibitem{b25} U. Zdun, R. Capilla, H. Tran, and O. Zimmermann. Sustainable architectural design decisions. \textit{IEEE Software}, 30(6), 2013.
\bibitem{b26} O. Kopp and A. Armbruster. Markdown architectural decision records: format and tool support. \textit{ZEUS Workshop}, 2018. \url{https://adr.github.io/madr/}
\bibitem{b27} O. Zimmermann. A definition of done for architectural decision making. \url{https://ozimmer.ch/practices/2020/05/22/ADDefinitionOfDone.html,} 2020.
\bibitem{b28} M. Fowler. Mocks aren't stubs. martinfowler.com, 2007.
\bibitem{b29} P. Hauer. Focus on integration tests instead of mock-based tests. phauer.com, 2019.
\bibitem{b30} M. Harman, J. Ritchey, I. Harper, S. Sengupta, K. Mao, A. Gulati, C. Foster, and H. Robert. Mutation-Guided LLM-based Test Generation at Meta. \textit{Proc. 33rd ACM Int. Conf. Foundations of Software Engineering (FSE Companion '25)}, Trondheim, Norway, 2025, pp. 180--191.
\bibitem{b31} J. C. Davis, P. C. Amusuo, T. Singla, B. Çakar, and K. A. Davis. Cheap code, costly judgment: A case study on governable agentic software engineering. arXiv:2607.01087, 2026.
\bibitem{b32} A. E. Hassan, H. Li, D. Lin, B. Adams, T.-H. Chen, Y. Kashiwa, and D. Qiu. Agentic software engineering: Foundational pillars and a research roadmap. arXiv:2509.06216, 2025.
\bibitem{b33} E. T. Barr, M. Harman, P. McMinn, M. Shahbaz, and S. Yoo. The oracle problem in software testing: A survey. \textit{IEEE Transactions on Software Engineering}, 41(5):507--525, 2015.
\bibitem{b34} L. Yang, C. Yang, S. Gao, W. Wang, B. Wang, Q. Zhu, X. Chu, J. Zhou, G. Liang, Q. Wang, and J. Chen. On the evaluation of large language models in unit test generation. \textit{Proc. ASE}, 2024.
\bibitem{b35} D. Taibi, P. Lago, and H. Muccini. Big questions on software architecture: report of the ICSE 2026 BoF on software architecture. arXiv:2609.04212, 2026.
\bibitem{b36} Z. Codabux and M. Galster. Reframing software architecture and design skills in the era of generative artificial intelligence. \textit{Proc. ICSE}, 2026.
\bibitem{b38} U. van Heesch. Dear researchers: The craft that I have practiced for the last twenty years no longer exists: What happens when AI agents become operational actors in software engineering? \textit{Journal of Systems and Software}, vol. 242, Art. no. 113000, 2026.
\bibitem{b39} B. Meyer. The ABC of software engineering. Blog post, 2013. \url{https://bertrandmeyer.com/2013/03/25/the-abc-of-software-engineering/}
\bibitem{b40} A. Cockburn. \textit{Agile Software Development: The Cooperative Game}, 2nd ed. Addison-Wesley, 2006.
\bibitem{b41} S. Baltes, F. Angermeir, C. Arora, et al. Guidelines for empirical studies in software engineering involving large language models. \textit{Empirical Software Engineering}, 2026. llm-guidelines.org; arXiv:2508.15503.
\bibitem{b42} M. Cohn. \textit{User Stories Applied: For Agile Software Development.} Addison-Wesley, 2004.
\bibitem{b43} B. Böckeler. Harness engineering for coding agent users. martinfowler.com, 2026. \url{https://martinfowler.com/articles/harness-engineering.html}
\bibitem{b44} N. D. Q. Bui. Building effective AI coding agents for the terminal: scaffolding, harness, context engineering, and lessons learned. \textit{arXiv}, 2026. arXiv:2603.05344.
\bibitem{b45} J. Becker, N. Rush, E. Barnes, and D. Rein. Measuring the impact of early-2025 AI on experienced open-source developer productivity. arXiv:2507.09089, 2025.
\bibitem{b46} J. T. Liang, C. Yang, and B. A. Myers. A large-scale survey on the usability of AI programming assistants: Successes and challenges. \textit{Proc. ICSE}, 2024.
\bibitem{b47} M. Galster, S. Mohsenimofidi, J. L. Lulla, M. A. Abubakar, C. Treude, and S. Baltes. Harness engineering for agentic AI coding tools: An exploratory study. 2026. arXiv:2602.14690.
\bibitem{b48} H. He, S. Agarwal, Y. Denisov-Blanch, P. Azaletskiy, S. Koyejo, and B. Vasilescu. AI writes faster than humans can review: A longitudinal study of an enterprise 2x mandate. arXiv:2607.01904, 2026.
\bibitem{b49} A. Simkute, L. Tankelevitch, V. Kewenig, A. E. Scott, A. Sellen, and S. Rintel. Ironies of generative AI: Understanding and mitigating productivity loss in human-AI interaction. \textit{International Journal of Human--Computer Interaction}, 41(5), 2025.
\bibitem{b50} H. Mozannar, G. Bansal, A. Fourney, and E. Horvitz. Reading between the lines: Modeling user behavior and costs in AI-assisted programming. \textit{Proc. CHI}, 2024.
\bibitem{b51} N. Perry, M. Srivastava, D. Kumar, and D. Boneh. Do users write more insecure code with AI assistants? \textit{Proc. ACM CCS}, 2023.
\bibitem{b52} K. Z. Cui, M. Demirer, S. Jaffe, L. Musolff, S. Peng, and T. Salz. The effects of generative AI on high-skilled work: Evidence from three field experiments with software developers. SSRN Working Paper, Aug. 2025. doi: 10.2139/ssrn.4945566.
\bibitem{b53} W. Takerngsaksiri, J. Pasuksmit, P. Thongtanunam, C. Tantithamthavorn, R. Zhang, F. Jiang, J. Li, E. Cook, K. Chen, and M. Wu. Human-in-the-loop software development agents. \textit{Proc. ICSE SEIP}, 2025.
\bibitem{b54} U. van Heesch, O. Zimmermann, and C. Kohls. What will remain human in software architecture? A focus group report. In \textit{Proc. 31st European Conference on Pattern Languages of Programs, People, and Practices (EuroPLoP 2026)}, 2026. arXiv:2609.30334.
\bibitem{b55} B. Flyvbjerg. Five misunderstandings about case-study research. \textit{Qualitative Inquiry}, 12(2):219--245, 2006.
\end{thebibliography}

\end{document}